# PolyLinkR: A linkage-sensitive gene set enrichment R package

Raymond Tobler[1,*], Angad Johar[1], Christian D. Huber[1], Yassine Souilmi[1,*]

[1]Australian Centre for Ancient DNA, University of Adelaide, North Terrace Campus, Adelaide, SA 5005, Australia
*Correspondence
RT raymond.tobler@adelaide.edu.au; YS yassine.souilmi@adelaide.edu.au

## Abstract

**Summary:** We introduce PolyLinkR, an R package for gene set enrichment analysis that implements a novel null-model algorithm that accounts for linkage disequilibrium between genes belonging to the same gene set – a potential cause of false positives that is not controlled for in similar tools. Our benchmarks show that PolyLinkR has improved performance compared to two similar tools, achieving comparable power to detect enriched gene sets while producing less than one falsely detected gene set on average, even at high genetic clustering levels and nominal false discovery rates of 20%.

**Availability and Implementation:** The R package is available under open-source MIT license https://github.com/ACAD-UofA/PolyLinkR.

**Contact:** RT raymond.tobler@adelaide.edu.au, YS yassine.souilmi@adelaide.edu.au

**Supplementary Information:** Supplementary data are available at *Bioinformatics* online.

## 1 Introduction

Gene set enrichment analysis (GSEA) is a widely used statistical approach for testing if specific classes of genes (defined by their biological function) are overrepresented amongst a larger set of genes that have been assayed in relation to a given phenotype (Hung *et al.*, 2012). Whereas conventional gene set enrichment tests are based on overrepresentation of pre-defined candidate genes, GSEA methods summarise information from all genes in a given gene set, making them particularly powerful for detecting sets of genes exhibiting small but consistent genetic signals that individually would not exceed standard significance thresholds (Tintle *et al.*, 2009). Such subtle signals are characteristic of polygenic selection – which may be a dominant mode of selection amongst humans and other species (Pritchard *et al.*, 2010) – whereby GSEA methods have become an increasingly popular means to detect biological pathways targeted by positive selection(Beichman *et al.*, 2019; Mayol *et al.*, 2020; Roux *et al.*, 2014).

Recently, the current authors developed a novel GSEA approach to identify biological pathways under positive selection in recent human history (Souilmi *et al.*, 2020). Other available enrichment tests either do not use all available genes to evaluate significance, or do not include a facility that explicitly corrects for the clustering of genes within gene sets. If left unchecked, gene clustering can inflate false discovery rates (FDR) in gene set enrichment studies – most notably, when positive selection targets a gene in linkage disequilibrium with functionally related genes. Here, we introduce PolyLinkR – a user-friendly R software implementation of our GSEA algorithm – and benchmark its performance against two popular gene set enrichment tools using simulated population genetic data. We show that PolyLinkR achieves comparable power to clustering-naive algorithms while significantly reducing the false positive rate associated with genomically clustered genes.

# 2 Methods

## 2.1 Algorithm

To reduce false positives due to gene clustering within biological pathways, we developed a novel algorithm that generates a null distribution that accounts for the spatial autocorrelation in genetic signals created by linkage disequilibrium (Souilmi *et al.*, 2020). Briefly, our permutation algorithm first arranges all gene scores linearly according to their chromosomal position, then randomly links all chromosomes into a single linear genome according to a permuted chromosome ordering. Reassignment of the scores to genes is performed by 'circularizing' (i.e. joining the terminal ends) the linear genome, which is then randomly rotated to create a unique remapping of the gene scores to the genes. The reassigned gene scores are summed within each gene set to generate a single null realization of each gene set score. Repeating this process a large number of times generates a null distribution of scores for each gene set, which is used to compute the associated *p*-value. Finally, multiple testing is controlled for by estimating *q*-values from the *p*-values of all tested gene sets (Storey, 2003). A full technical description of the algorithm is provided in the Supplementary Materials.

## 2.2 Simulating genomic clustering

We evaluated whether the PolyLinkR permutation algorithm was able to effectively minimise the number of false positives due to gene clustering while maintaining comparable power to a widely-used, clustering-naive GSEA method, PolySel (Daub *et al.*, 2013). For each gene set with *n* genes, PolySel creates a null distribution by summing scores from *n* genes randomly chosen from across the genome. We also compared both GSA approaches to another popular Gene Ontology enrichment package, GOfuncR (Grote, 2019). Unlike PolyLink and PolySel, GOfuncR requires predefined candidate genes or genomic regions; however, it is one of the few available enrichment methods known to the authors that also implements a null model that controls for gene clustering (see Supplementary Materials).

To measure the performance of each tool, we simulated a series of genomic datasets, each comprising 22 chromosomes that match human autosomes in length, which were seeded with normally distributed scores. These scores were simulated at four different levels of genomic autocorrelation (i.e. $1x10^5$bp, $5x10^5$bp, $1x10^6$bp, $5x10^6$bp) – i.e. covariance in neighbouring genome scores caused by linkage disequilibrium and other evolutionary processes – which equate to moderate to high levels of autocorrelation observed in human population genomic datasets (see Figures S1, S2, Supplementary Materials). We combined these simulated genomic datasets with ~3,500 annotated gene sets to create inputs for PolyLinkR, PolySel, and GOfuncR and estimated the number of false positives and power for each method of under a range of different false discovery rates (FDR) commonly used in enrichment studies (further details are provided in the Supplementary Materials).

## 3 Results

Amongst the methods tested here, PolyLinkR offers the best trade-off between detection power and false positives across all levels of genomic autocorrelation – detecting >60% of selected gene sets with less than one false positive on average, even at a stringent FDR of 2% (Figures 1, S3, S4) – and showed considerably less detection bias for highly clustered gene sets than the other two tested methods (Figure S2). While PolySel has up to 20% more power than PolyLinkR at relaxed FDRs of 10% and 20%, this increase in power comes with an average of 10 to 30 times more false-positive gene sets than PolyLinkR. In contrast, GOfuncR had few false positives on average at low autocorrelation levels, but false positives became more frequent as autocorrelation increased and detection power was consistently lower than PolyLinkR. While PolyLinkR is slower than PolySel when using the same number of iterations – testing ~3,000 gene sets and 20,000 genes takes ~52 seconds on a single 3.5khz Intel i7® processor for 10,000 iterations of the PolyLinkR permutation algorithm, vs only ~8 seconds for PolySel (computation time scales linearly with the number of iterations for both methods) – it is sufficiently fast that runtime is unlikely to be a problem for most datasets.

## 4 Conclusion

PolyLinkR is a user-friendly gene set enrichment approach that explicitly accounts for gene clustering, having fewer false-positive gene sets and comparable power relative to similar tools. PolyLinkR can be used for any organism, genetic assay (e.g. data from gene expression, population genetic, or other omics assays), or gene set source (e.g. any annotated set of genes), and is available as a platform-agnostic R package (https://github.com/ACAD-UofA/PolyLinkR).

## Acknowledgements

The authors thank Alan Cooper for naming the method. This work was supported by the Australian Research Council [DE190101069 to RT, DE180100883 to CDH; YS and AJ were funded by DP190103705 and FL140100260, respectively].

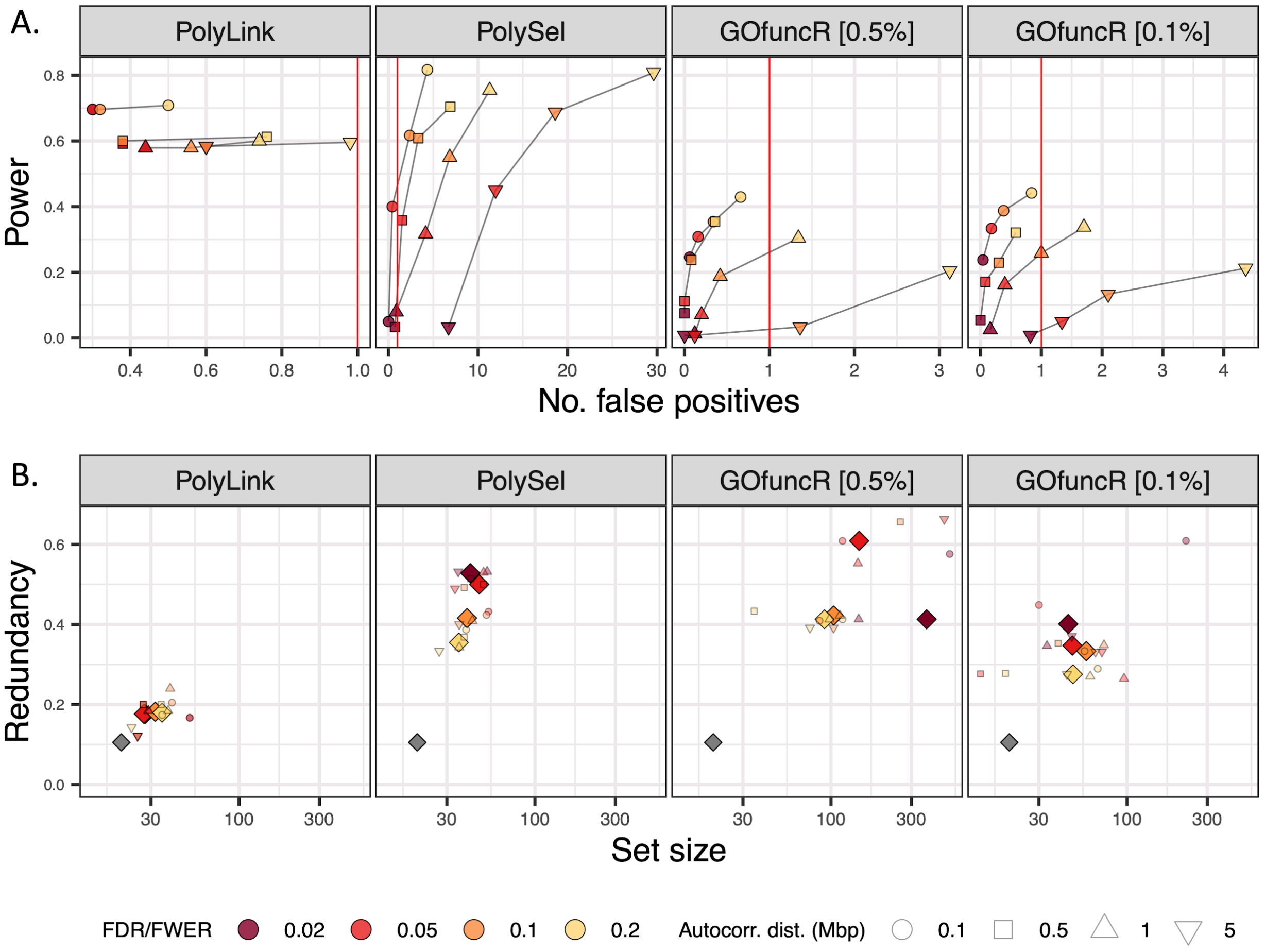

**Fig. 1. Benchmarking PolyLinkR performance. (A)** Statistical power evaluated against the number of false positives detected for three different gene set enrichment methods (panels; GOfuncR measured at 2 different candidate gene thresholds). **(B)** Average set size and gene clustering (measured as redundancy; equals 0 no gene lies within 1cM of another gene from the same gene set, and approaches 1 when all genes in the same set sit within 1cM of each other, see Supplementary Methods) amongst significantly enriched gene sets. For both panels, values were quantified at four levels of genomic spatial autocorrelation (denoted by different shapes) and false discovery rate (FDR; PolyLinkR and PolySel) or family-wise error rate (FWER; GOfuncR) thresholds (denoted by different colours). The red line in panel A demarcates 1 false-positive result, coloured diamonds in panel 2 denote the median redundancy/set size for each FDR/FWER level, with grey diamonds showing the median values for all gene sets.

Supplementary Materials

# PolyLinkR: A linkage-sensitive gene set enrichment R package

Raymond Tobler[1,*], Angad Johar[1], Christian D. Huber[1], Yassine Souilmi[1,*]

## 1. PolyLinkR null distribution generation procedure

The PolyLinkR method uses a gene set enrichment approach (GSEA) to test if any predefined gene sets have a significant excess of signals arising from specific biological assay; e.g. highly expressed genes in an RNAseq assay (Subramanian *et al.*, 2005) or highly divergent genes in a population genomic scan of multiple populations (Daub *et al.*, 2013). Following recent approaches (Daub *et al.*, 2013), the core PolyLinkR algorithm contrasts the sum of all gene scores within a given gene set to a null distribution of gene scores, which has been shown to be more powerful than other commonly used GSEA statistics (Tintle *et al.*, 2009). The key advantage of PolyLinkR is that rather than generating the null distribution for each gene set by repeatedly summing scores from the same number of genes randomly sampled from across the genome – which does not account for gene clustering within gene sets – PolyLinkR uses a permutation procedure that preserves the genomic linkage inherent in the genome, as well as the gene clustering specific to each gene set. To do so, the PolyLinkR permutation method incorporates the two following steps:

1. Gene scores are arranged linearly in each chromosome according to their chromosomal position, and chromosomes are randomly joined together (preserving 5' to 3' orientation). Specifically, the random joining of chromosomes is achieved by permuting the set of chromosome labels and then connecting the chromosomes according to this permutation.

2. To maximize the chance that unique gene assignments are achieved during each iteration of the randomization procedure, we perform a further 'rotation' of the gene scores subsequent to joining the chromosomes in step 1. This is achieved by assigning a number to each gene according to their linear position following the chromosome joining step, then randomly selecting a single number from this range. Assuming that there are $N$ genes, if the randomly drawn number is $i$ (bounded by 2 and $N$), this results in a unique mapping where the gene score initially in position $x$ is re-assigned to position 1, the score initially in position $i$+1 is shifted to position 2, and so on until the score initially in position $N$ takes position $N$-$i$+1; similarly, the gene score initially in position 1 is now re-assigned to position $N$-$i$+2, the score initially in position 2 goes to position $N$-$i$+3 and so on until the score initially in position $i$-1 takes position $N$.

The total number of possible unique mappings using this process is $C$! x $N$-1, where $C$ is the number of chromosomes being tested. For the human genome with 23 chromosomes and >20,000 currently described genes, there are at least 23! x 20,000 or $>10^{25}$ unique mappings of gene scores to genes using this permutation procedure. The lower bound of the number of unique mappings is achieved for species with a genome comprising a single chromosome, where the number of permutations is simply the number of genes (minus one).

## 2. Simulating autocorrelated genetic values

Two sets of population genetic datasets were simulated to benchmark PolyLinkR performance: a set of neutral simulations were used for false-positive quantification, and another set of simulations that included selection on the genes in a prespecified gene set that was used to estimate power (see *Estimating power*). For both sets of simulations, we used the R package RandomFields (Schlather *et al.*, 2015, 2020) to generate autocorrelated gene scores for each of the simulations used in this study. Following previous results (Hofer *et al.*, 2012) and our own analyses of human population genetic datasets (see details in the following paragraph), we modeled gene scores as standard random normally distributed values (i.e. *Z* scores), using the RandomField function `RPgauss` to generate standard normal values every 1kbp across 22 separate chromosomes (each simulated chromosome replicating one of the human autosomes in length). Spatial autocorrelation was modelled using an exponential covariance function, `RPexp`, with variance set to 1 and scale modelled at four separate levels – $1x10^5$bp, $5x10^5$bp, $1x10^6$bp, and $5x10^6$bp – generating moderate to high levels of genomic autocorrelation that is consistent with levels observed in human population genetic datasets.

To determine the levels of genomic autocorrelation to use in our simulations, we used the `RMfit` function from the RandomFields R package to estimate the scale and variance parameters amongst 12 different population genetic summary statistics computed on the CEU 1000 genomes dataset, which were obtained from the Human Selection Database (Pybus *et al.*, 2014); http://hsb.upf.edu/hsb_data/positive_selection_NAR2013/?C=S;O=A). Each of these statistics was evaluated every 3kbp (in 30kbp-wide windows), which closely matched the density of values in our own simulated datasets (1kbp). Most of these summary statistics approximate a standard normal distribution (note that we $log_{10}$-transformed the EHHAv statistic), though several statistics had one heavy tail that is expected when a non-neutral process, such as positive selection, has impacted allele frequencies in the tested population (Figure S1).

To avoid the influence of outlier values on our estimates – which could indicate genomic regions that have been influenced by selection – we generated a list of 10Mbp genomic windows where the observed number of values in either the upper and lower 2.5% tails did not exceed the expected number by more than 5% (assuming that scores follow a standard normal distribution). We applied the `RMfit` function to 30 of the 10Mbp windows from this list, taking 3 windows from each of the first 10 chromosomes. Notably, the `RMfit` function provided reasonable estimates of the scale and variance parameters for simulated 10Mbp genomic regions with *Z* scores distributed every 3kbp (windows simulated using `RPgauss` and `RPexp`) when taking mean and median estimates across several simulations (Figure S2C). With the exception of a single statistic (EHHAv), the median estimated autocorrelation values for the 12 different summary statistics fell between ~50kbp to ~200kbp (Figure S1). Based on these results, we chose autocorrelation levels ranging between 100kbp and 5Mbp for our simulations, thereby facilitating the evaluation of the performance of PolyLinkR and related software under moderate to high levels of genomic autocorrelation.

## 3. False-positive gene set estimation

We used the simulated population genomic datasets described in Supplementary Materials section 2 to estimate the number of false-positive gene sets arising from PolyLink and another GSEA method, PolySel, along with a widely used GO enrichment tool, GOfuncR.

The PolyLinkR and PolySel procedures both require that the simulated genetic scores are assigned to genes. Accordingly, we assigned the simulated *Z* scores to ~20,000 separate annotated human genes from the ENSEMBL database (genome reference version: GRCH37; Kinsella *et al.*, 2011), which was accessed using the R bioMart package (version 2.36.1; Durinck *et al.*, 2009). We only used genes that had specific protein or RNA based annotation (in the bioMart transcript_biotype field) and which were also annotated in the NCBI database (ftp://ftp.ncbi.nih.gov/gene/DATA/GENE_INFO/Mammalia/Homo_sapiens.gene_info.gz), resulting in 19,603 genes in total. To assign *Z* scores to these genes, we extended the boundaries by 50kbp on either side of each gene in order to capture cis-regulatory regions, then took the maximum *Z* score overlapping this interval to represent each gene. Because this results in larger genes having higher scores on average, we used the non-parametric correction, $Z_i = 0.675 \times y_i/\text{median}(|y|)$, where $y_i$ is the score of gene *i* (Iglewicz and Hoaglin, 1993), to compute a set of standardised scores corrected for gene length.

After generating the set of gene scores, we combined these with ~3,500 gene sets from the Gene Ontology database (Ashburner *et al.*, 2000; The Gene Ontology Consortium, 2019), which was also accessed with the R bioMart package (Durinck *et al.*, 2009). Of the >6,000 gene sets on the database, we removed all sets with fewer than 10 genes or more than 1000 genes, resulting in 3,458 gene sets in total. This set of gene scores and gene sets were passed to PolyLinkR and PolySel to calculate *p*-values for each gene set. We ran each method on 50 different simulated datasets, using 10,000 iterations to estimate the null distribution in each instance. To control for false positives due to multiple testing on large numbers of gene sets, we applied the *q*-value correction (Storey, 2003) to the complete set of *p*-values. The *q*-value is a Bayesian posterior estimate of the *p*-value that accounts for the expected inflation of false positives due to multiple testing, whereby a *q*-value of 0.01 implies an FDR of 1%. We calculated the number of false-positive gene sets at *q*-values of 0.02, 0.05, 0.10, and 0.20, ranging from stringent to more moderate FDRs that are often used in enrichment studies.

For GOfuncR, the inbuilt method used to control for gene clustering requires characterising genomic regions into either selected or neutral regions, rather than assigning scores to genes. To determine selected regions, we initially identified outlier *Z* scores that exceeded a specific threshold – either 0.99 and 0.999 standard normal quantiles, or $Z > 2.33$ and $Z > 3.09$, respectively – and then combined together any sets of outlier scores that were situated within the autocorrelation distance used to generate the simulated dataset (i.e. $1\text{x}10^5$bp, $5\text{x}10^5$bp, $1\text{x}10^6$bp, or $5\text{x}10^6$bp). The resulting boundaries of these combined outlier scores were classified as the selected regions, and the remaining genomic regions as neutral regions. These genomic regions were passed to GOfuncR enrichment test function, `go_enrich`, setting the `regions` flag to TRUE and performing the recommended 1,000 iterations. Rather than estimating an FDR like PolyLinkR and PolySel, GOfuncR uses the family-wise error rate (FWER; the estimated probability of having at least one false positive below a given threshold) to correct for multiple testing. Hence we used 0.02, 0.05, 0.10, and 0.20 FWER values as the thresholds to determine false positive gene sets for GOfuncR.

## 4. Estimating power

To quantify the detection power of the PolyLinkR, PolySel, and GOfuncR we simulated population genetic datasets that also included a single selected gene set. For all genes from a predefined selected gene set, we added a randomly chosen positive Z score (i.e. a random draw from a half-normal distribution) to each of the original *Z* scores situated inside the boundaries of each gene. Because the sum of two random normal variables is also normally distributed (with the mean and standard deviation equal to the sum of each respective parameter from the two distributions), this resulted in all scores situated within the boundaries of selected genes being normally distributed with mean of $\sqrt{2/\pi}$ and standard deviation of $1 + (1 + 2/\pi) = 2 + 2/\pi$ (Leone *et al.*, 1961). Ten simulations were performed in this manner for each of 24 different gene sets that were chosen to span a wide range of gene set sizes and clustering levels, with three gene sets being randomly sampled for each of eight different combinations of redundancy level (<10%, 20-40%, and >50%; see *Quantifying gene clustering within gene sets*) and gene set sizes (10 to 50 genes, 51 to 150 genes, and 151 to 1000 genes) – note that there were no gene sets that had more than 150 genes with less than 10% redundancy (see Table S1 for list of tested gene sets).

For each simulation, the scores were appropriately processed and passed to each of the three methods for analysis using the same parameters and gene sets outlined in Supplementary Materials section 3. Power was computed at the four different FDR/FWER thresholds by measuring the proportion of times that each of the prespecified gene sets had a *p*-value lower than this threshold after adjusting for multiple testing.

## 5. Quantifying gene clustering within gene sets

To quantify the amount of gene clustering in each gene set, we computed a simple 'redundancy' index that ranged between 0 (minimum clustering, no redundant genes) and 1 (maximum clustering, all but one gene is redundant). For each gene set we ordered all genes according to their chromosomal position. Starting from the gene located in the most upstream position in each chromosome, i.e. the 'focal' gene, we successively merged any neighbouring gene that was within 1cM the focal gene into a 'fictive gene' (Daub *et al.*, 2013), with genetic distances defined using the sex-averaged genetic map obtained from Bhérer and colleagues (Bhérer *et al.*, 2017). The first gene not lying within 1cM of the focal gene becomes the new focal gene and this process is repeated for all genes in the gene set on each chromosome.

We applied this process to each of the ~3,500 gene sets to produce a list of fictive genes for each set. Because the number of fictive genes is necessarily equal to or less than the number of original genes for any given gene set, taking the ratio of number of fictive genes, $N_f$, to original genes, $N_o$, returns a value between $1/N_o$ and 1 – the lower bound occuring when all genes in a gene set are within 1cM from the first gene (i.e. creating a single fictive gene), and the upper bound arising when no gene in the gene set sits within 1cM of another gene in that set (i.e. $N_f = N_o$). The redundancy index is computed as $1 - N_f / N_o$, whereby values of 0 indicate no clustering and those approaching 1 indicate maximal clustering for a particular gene set.

## 6. R package

The PolyLinkR R package is available at https://github.com/ACAD-UofA/PolyLinkR. It contains three functions – a data input function (`ReadSetObjTables`), a function that executes the enrichment test (`polylinkr`), and a plotting function (`plot_polyviolinr`) that produces a violin plot of the score distribution for the enriched gene sets. Details about each function can be accessed by querying the relevant function name (e.g. ?`polylinkr`) in an R session. An input dataset is also included. The input format and functionality of PolyLinkR is modeled on the popular GSEA method for population genetic datasets, PolySel (https://github.com/CMPG/polysel). PolyLinkR is designed to be used with data from any organism, genetic assay (e.g. data from gene expression, population genetic, or other omics assays), or gene set source (e.g. from any source of annotated gene sets; e.g. BIOCYC (Romero *et al.*, 2005), KEGG (Kanehisa and Goto, 2000), PID (Schaefer *et al.*, 2009), and REACTOME (Schaefer *et al.*, 2009; Fabregat *et al.*, 2018)).

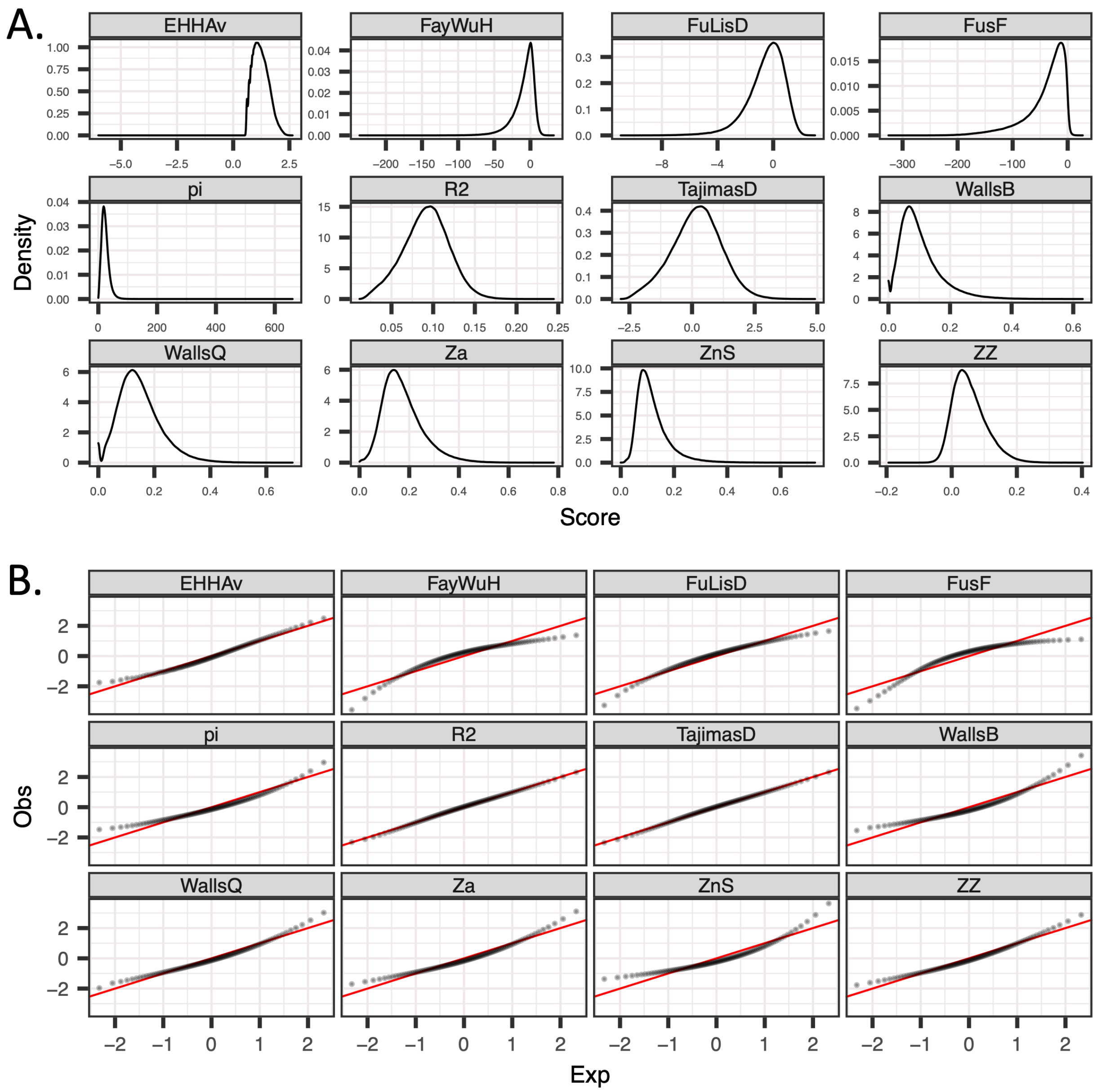

**Figure S1. Human population genetic summary statistic distributions. (A)** Distribution of 12 summary statistics obtained from the Human Selection Database (see (Pybus *et al.*, 2014) for details on each summary statistic). **(B)** QQ plots of each summary statistics showing that they approximate a standard normal distribution with heavy tails, potentially indicative of positive selection (note that EHHAv has been $log_{10}$ transformed).

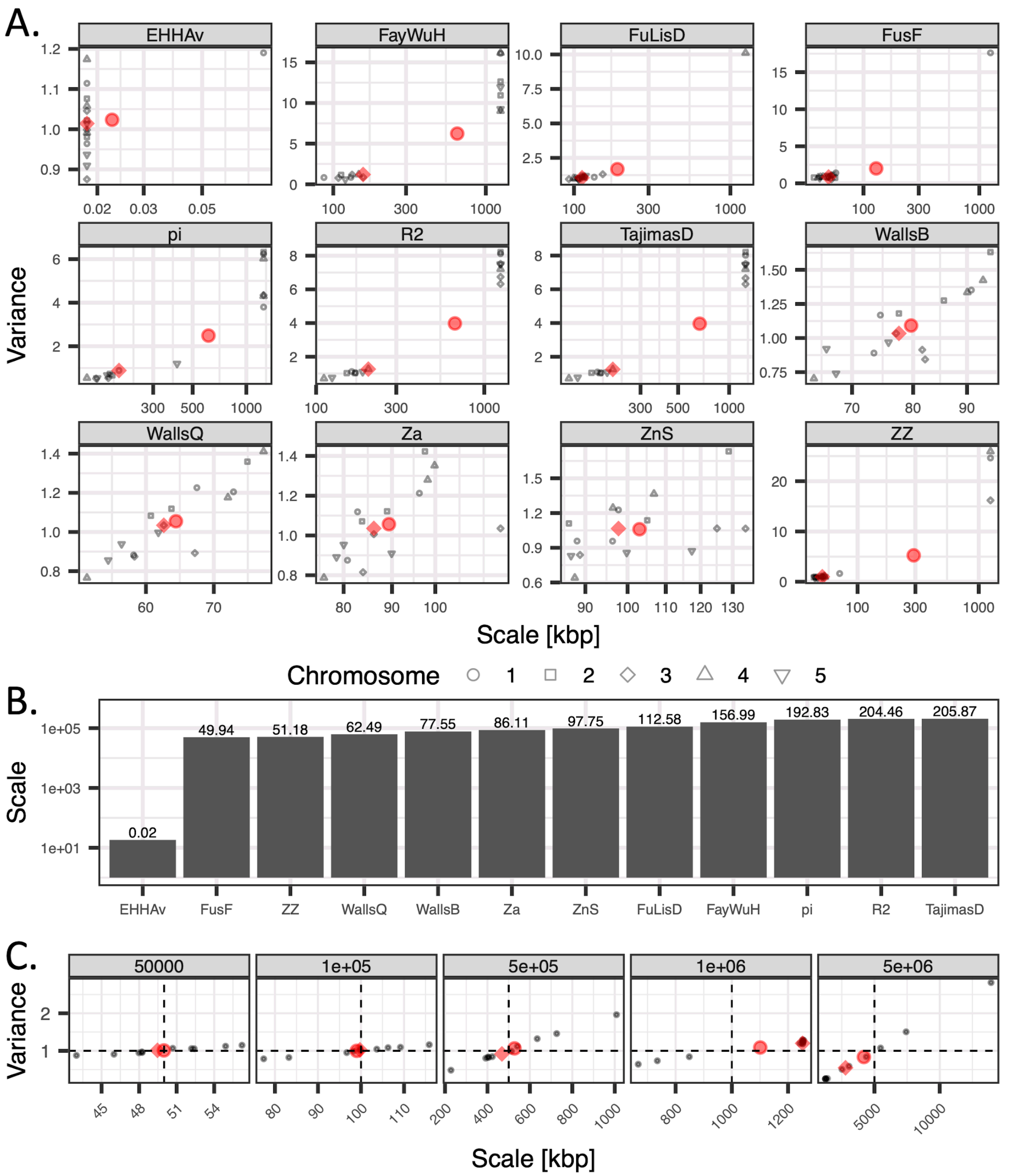

**Figure S2. Inferring autocorrelation levels from human population genetic summary statistics. (A)** Estimation of the autocorrelation (scale, x-axis) and variation (y-axis) amongst 12 summary statistics (facets) obtained from the Human Selection Database. Each point indicates the values obtained from one of the 15 evaluated 10Mb genomic regions. The mean and median estimates are shown as red circles and diamonds, respectively. Parameter inference used the `RMfit` function from the RandomFields R package. **(B)** The median estimated autocorrelation level for each method (value on top of each bar, in units of 1kbp). These values were used to establish the scale parameter for the simulated data in the present study. **(C)** The performance of `RMfit` to estimate scale and variance from simulated datasets with similar properties as the summary statistic data from the Human Selection Database. Simulations were performed at four autocorrelation levels used in this study and also a lower value, 50kbp (see facet labels). Dashed lines indicate parameter expectations.

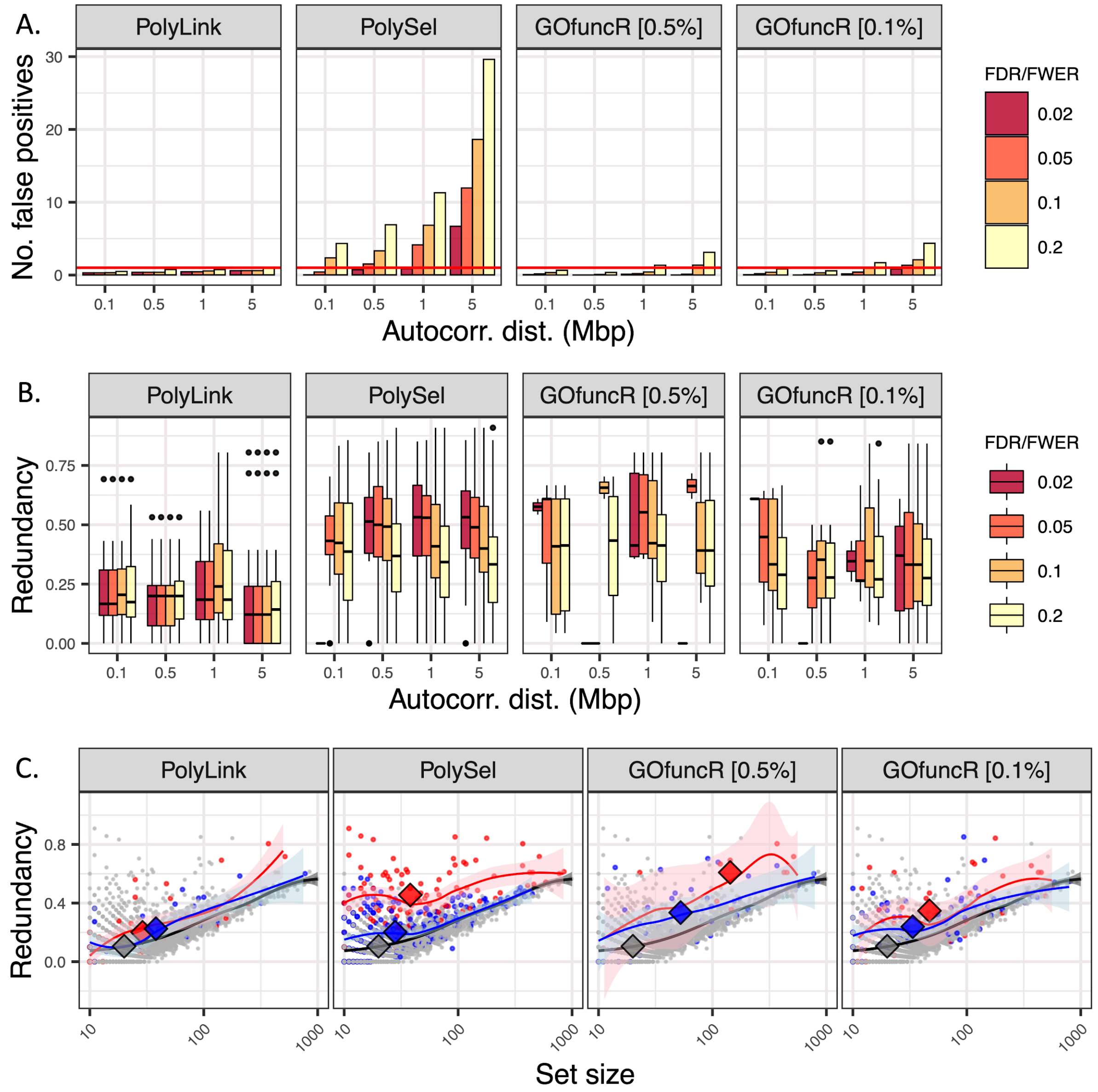

**Figure S3. Robustness of enrichment methods to gene clustering. (A)** The number of false positive gene sets detected for two different gene set enrichment methods (PolyLinkR and PolySel) along with a standard gene ontology enrichment method (GOfuncR) measured at 2 different candidate gene thresholds (square brackets in facet label). False positives were quantified at four different levels of genomic spatial autocorrelation (x-axis) and false discovery rates (FDR; PolyLinkR and PolySel) or family-wise error rates (FWER; GOfuncR) thresholds (denoted by different colours). The horizontal red line indicates one false positive. **(B)** Same as panel A, but showing the level of redundancy in each gene set for each level of spatial autocorrelation (see explanation in Fig 1B and Supplementary Materials). **(C)** The level of gene clustering in each gene set shows a log-linear relationship with the number of genes in the set (Set size). Gene sets detected as significant at FDR/FWER < 0.05 or 0.20 in each method are shown as red or blue points, respectively, and all other gene sets are shown as grey points, with local regression lines fitted for false positives (red = FDR/FWER < 0.05; blue = FDR/FWER < 0.20) and all gene sets (black). Detected gene sets in PolyLinkR have median redundancy (red or blue diamonds) that is closer to the expected redundancy levels (grey diamonds) relative to the other tested methods, indicating that PolyLinkR has less detection bias from gene clustering, particularly at more stringent FDRs.

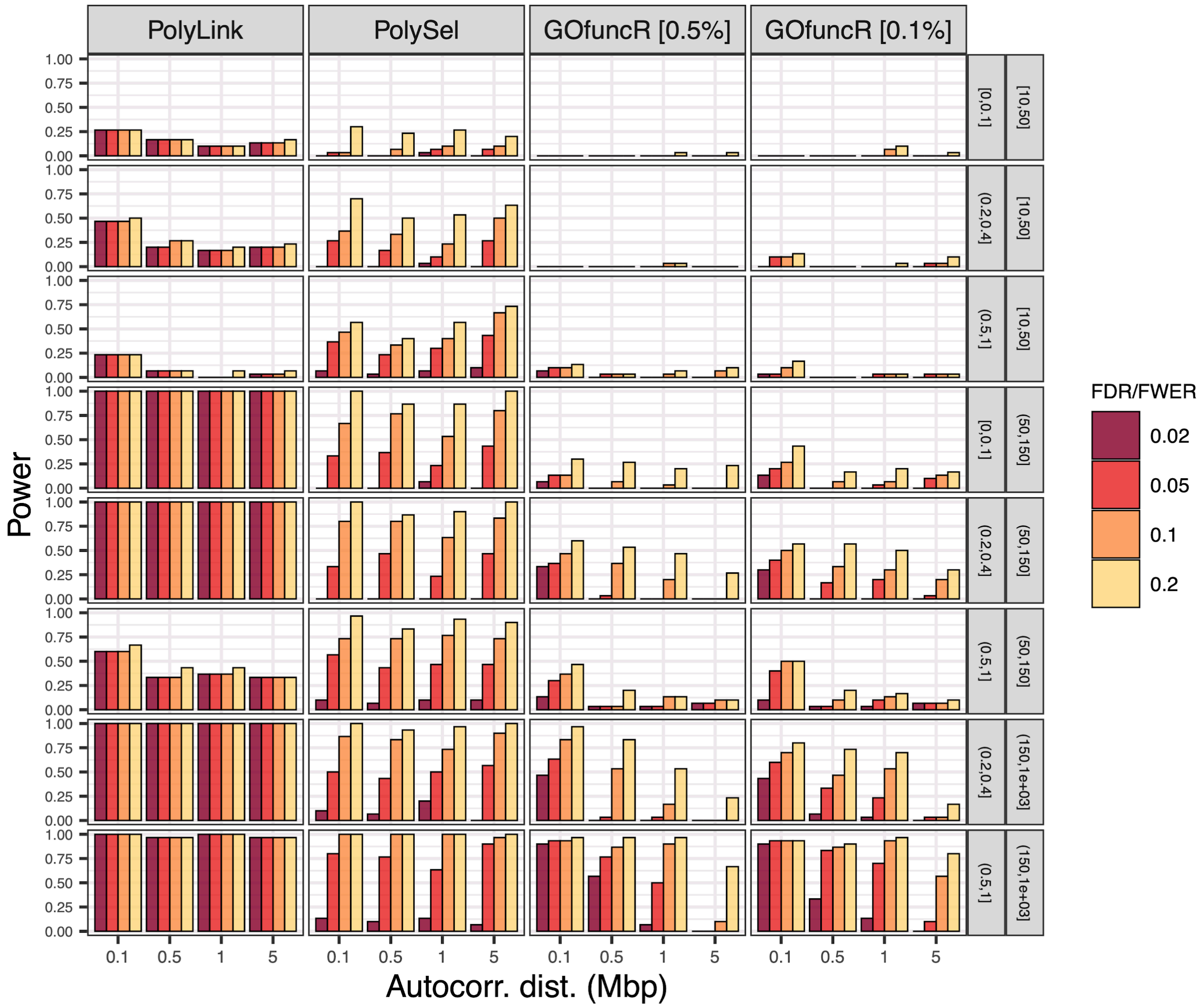

**Figure S4. Power under different levels of gene clustering.** Same as Figure S1A, but showing the average power to detect a gene set with multiple selected genes across four levels of spatial autocorrelation and eight different combinations of redundancy level (<10%, 20-40%, and >50%) and gene set sizes (10 to 50 genes, 51 to 150 genes, and 151 to 1000 genes) – note that there were no gene sets that had more than 150 genes with less than 10% redundancy. For all methods, power tends to increase as set sizes get larger, a consequence of the larger number of genes that are coming under selection. Power also has a positive association with clustering level for PolySel and GOfuncR, though not for PolyLinkR, particularly when gene sets have less than 150 genes. This suggests that the PolyLinkR algorithm is more punitive for highly clustered gene sets of small to moderate sizes.

**Table S1. Gene sets used in simulations for power estimation.** The name of each gene set used for power estimation (column 1), including Gene Ontology ID (column 2), biological domain (column 3), number of genes in gene set (column 4), number of fictive genes in gene set (column 5), Redundancy (column 6), set size class (column 7), and clustering level class (column 8).

| Name | GO_ID | Domain | #Genes | #Fictive | Red. | Set size | Cl. level |
|---|---|---|---|---|---|---|---|
| protein stabilization | GO:0050821 | biological_process | 156 | 118 | 24.359% | (150,1e+03] | (0.2,0.4] |
| nuclear chromatin | GO:0000790 | cellular_component | 162 | 113 | 30.247% | (150,1e+03] | (0.2,0.4] |
| neuron projection | GO:0043005 | cellular_component | 164 | 112 | 31.707% | (150,1e+03] | (0.2,0.4] |
| micro-ribonucleoprotein complex | GO:0035068 | cellular_component | 425 | 166 | 60.941% | (150,1e+03] | (0.5,1] |
| integral component of plasma membrane | GO:0005887 | cellular_component | 833 | 377 | 54.742% | (150,1e+03] | (0.5,1] |
| mitochondrion | GO:0005739 | cellular_component | 992 | 431 | 56.552% | (150,1e+03] | (0.5,1] |
| protein homotetramerization | GO:0051289 | biological_process | 83 | 57 | 31.325% | (50,150] | (0.2,0.4] |
| Z disc | GO:0030018 | cellular_component | 101 | 77 | 23.762% | (50,150] | (0.2,0.4] |
| extracellular matrix | GO:0031012 | cellular_component | 103 | 68 | 33.981% | (50,150] | (0.2,0.4] |
| nucleosome assembly | GO:0006334 | biological_process | 84 | 33 | 60.714% | (50,150] | (0.5,1] |
| negative regulation of endopeptidase activity | GO:0010951 | biological_process | 97 | 48 | 50.515% | (50,150] | (0.5,1] |
| homophilic cell adhesion via plasma membrane adhesion molecules | GO:0007156 | biological_process | 117 | 39 | 66.667% | (50,150] | (0.5,1] |
| cytoplasmic stress granule | GO:0010494 | cellular_component | 51 | 46 | 9.804% | (50,150] | [0,0.1] |
| cellular response to transforming growth factor beta stimulus | GO:0071560 | biological_process | 55 | 51 | 7.273% | (50,150] | [0,0.1] |
| motile cilium | GO:0031514 | cellular_component | 72 | 65 | 9.722% | (50,150] | [0,0.1] |
| 3'-phosphoadenosine 5'-phosphosulfate metabolic process | GO:0050427 | biological_process | 18 | 12 | 33.333% | [10,50] | (0.2,0.4] |
| connexin complex | GO:0005922 | cellular_component | 19 | 14 | 26.316% | [10,50] | (0.2,0.4] |
| adherens junction organization | GO:0034332 | biological_process | 30 | 23 | 23.333% | [10,50] | (0.2,0.4] |
| taste receptor activity | GO:0008527 | molecular_function | 14 | 6 | 57.143% | [10,50] | (0.5,1] |
| calcium-dependent cell-cell adhesion via plasma membrane cell adhesion molecules | GO:0016339 | biological_process | 25 | 11 | 56.000% | [10,50] | (0.5,1] |
| DNA replication-dependent nucleosome assembly | GO:0006335 | biological_process | 32 | 13 | 59.375% | [10,50] | (0.5,1] |
| Gemini of coiled bodies | GO:0097504 | cellular_component | 10 | 9 | 10.000% | [10,50] | [0,0.1] |
| positive regulation of TORC1 signaling | GO:1904263 | biological_process | 10 | 10 | 0.000% | [10,50] | [0,0.1] |
| cellular response to peptide | GO:1901653 | biological_process | 13 | 13 | 0.000% | [10,50] | [0,0.1] |